\documentstyle[aps,prl,twocolumn,tighten,epsf]{revtex}
\def\addcontentsline#1#2#3{\relax}
\begin{document}

\title{Detecting Gapless Excitations above Ferromagnetic Domain Walls}

\author{Tohru Koma$^1$$^*$ and Masanori Yamanaka$^2$$^{\dagger}$}

\address{$^1$Department of Physics, Gakushuin University, 
Mejiro 1-5-1, Toshima-ku, Tokyo 171-8588, Japan}

\address{$^2$Department of Applied Physics, Science University of Tokyo, 
Kagurasaka 1-3, Shinjuku-ku, Tokyo 162-8601, Japan}

\maketitle

\begin{abstract}
In a two or three dimensional ferromagnetic XXZ model,
a low energy excitation mode above a magnetic domain wall is gapless,
whereas all of the usual spin wave excitations moving 
around the whole crystal are gapful. 
Although this surprising fact was already proved in a mathematically 
rigorous manner, 
the gapless excitations have not yet been detected experimentally.
For this issue, we show theoretically 
that the gapless excitations appear as the dynamical fluctuations 
of the experimental observable, magnetoresistance,
in a ferromagnetic wire. 
We also discuss other methods (e.g., ferromagnetic resonance and 
neutron scattering) to detect the gapless excitations experimentally. 
\end{abstract}

\pacs{72.15.-v, 72.10.-d, 73.40.-c, 76.50.+g,}

In ferromagnetic spin systems, 
the low energy excitations above the ground states 
can be described by the conventional spin wave theory 
under the assumption that no magnetic domain structure 
affects the low energy spectrum. 
However, this assumption is not necessarily valid. In fact, 
it was recently proved \cite{{REFKN2},{REFmatsui2}} 
in a mathematically rigorous manner that, 
in a two or higher dimensional ferromagnetic XXZ model with 
arbitrary spin $S$, there exists an excitation above 
a magnetic domain wall such that the excitation is totally 
different from those by the conventional spin wave theory. 
More precisely, the excitation mode above a magnetic domain wall 
is gapless, whereas all of the usual spin wave excitations moving 
around the whole crystal are gapful. 
Namely all of the gapless excitations are confined in the narrow region 
along a magnetic domain wall. 

This locality makes it very difficult to detect the gapless 
excitations in experiments. Thus the gapless excitations 
have not been found experimentally so far. 
However, quite recent technology has made it possible to create 
only one single domain wall in a ferromagnet with a very small size. 
For example, the resistance contributions due to a single 
ferromagnetic domain wall were actually measured in a magnetoresistance 
experiment for a metallic wire \cite{Otani}. 
An experimentally important question is whether 
the gapless excitations can be actually detected or not. 
We can list at least two experimental methods which have 
a possibility to detect directly the spectrum of the gapless excitations. 
These are ferromagnetic resonance and neutron scattering. 
As is well known, by these methods, 
one can get the Fourier transform of the spin-spin correlation 
which is directly related to the low energy excitations 
above the ground states of the system at a very low temperature. 
Clearly the dominant part of the Fourier spectrum 
consists of the contributions from the usual spin wave excitations. 
In order to obtain only the spectrum from the gapless excitations, 
the small contributions must be separated from the dominant contributions 
due to the usual spin wave excitations. 
Since we can expect that this detection will succeed by 
overcoming the technical problem in the near future, 
we will briefly discuss this issue in this Letter. 

The main aim of this Letter is to propose another experimental method 
to find the evidence of the gapless excitations 
above a ferromagnetic domain wall. 
This is a method to detect the dynamical fluctuations 
in the magnetoresistance due to the gapless excitations. 
Although the method is indirect in comparison to the above direct methods, 
the detection of the dynamical fluctuations due to the gapless excitations 
is itself very interesting and challenging problem 
from both theoretical and experimental points of view. 
In order to get the dynamical fluctuations of the resistance, 
we calculate the transmission coefficient for a conduction electron through 
the ferromagnetic domain wall with gapless excitations derived by 
deforming the spin configuration of the domain wall. 
It turns out that the transmission coefficient 
depends on the detailed structures of gapless excitations, in particular, 
the position of a kink excitation above the domain wall. 
This implies that the dynamical fluctuations of the low energy gapless mode 
appear through the transmission coefficient. According to the Landauer formula 
for a quantum wire, the conductance is proportional to the transmission 
coefficient through the wire. Thus the dynamical fluctuations 
of the low energy gapless mode appear as the resistance fluctuations. 

In order to calculate the transmission coefficient through the domain wall, 
we use a Schr\"odinger equation \cite{CF} with the effective domain wall 
potential for the conduction electron. The equation is microscopically 
derived from \cite{REFYK} a ferromagnetic XXZ-Kondo model 
in a two-dimensional box $\Omega:=[-L_x/2,L_x/2]\times[-L_y/2,L_y/2]$. 
(In the same way, we can treat the same system in three or higher dimensions.) 
The total Hamiltonian consists of three terms as 
\begin{equation}
H=H_{\rm el}+H_{\rm dw}+H_{{\rm el}-{\rm dw}}. 
\label{eq:hamiltonian}
\end{equation}
The kinetic term $H_{\rm el}$ for the conduction electron is given by 
\begin{equation}
H_{\rm el}=-\frac{\hbar^2}{2\mu}\triangle
=-\frac{\hbar^2}{2\mu} 
\left(\frac{d^2}{dx^2}+\frac{d^2}{dy^2}\right) 
\end{equation}
with the mass $\mu$ of the electron. 
The ferromagnetic XXZ Hamiltonian $H_{\rm dw}$ for the domain wall 
\cite{REFASW} is 
\begin{eqnarray}
H_{\rm dw}&=&-J\sum_{\langle {\bf a},{\bf b}\rangle}
\left[S_{\bf a}^{(1)}S_{\bf b}^{(1)}+S_{\bf a}^{(2)}S_{\bf b}^{(2)}
+\Delta S_{\bf a}^{(3)}S_{\bf b}^{(3)}\right]\nonumber\\
&-&J\sqrt{\Delta^2-1}\left[\sum_{{\bf a}\in B_+}
S_{\bf a}^{(3)}-\sum_{{\bf b}\in B_-}S_{\bf b}^{(3)}\right]
\label{hamXXZ} 
\end{eqnarray}
with the nearest neighbor spin-spin interactions 
with the exchange integral $J>0$ and the anisotropy $\Delta>1$.
Here ${\bf S}_{\bf a}=(S^{(1)}_{\bf a},S^{(2)}_{\bf a},S^{(3)}_{\bf a})$ 
is the spin-1/2 operator at the site ${\bf a}=(a_x,a_y)$ in 
the two-dimensional diagonal lattice \cite{diagonal}
$\Lambda=\{(ma,na)\in\Omega|\ \mbox{two integers $m,n$ 
satisfy $m+n=$even}\}$ 
with a lattice constant $a$. (See Fig.~1.) For simplicity, 
we take $4Ma=L_x$, $2(2N+1)a=L_y$ with positive integers $M,N$. 
In order to make a single domain wall ground state, 
we have applied the boundary fields $\pm J\sqrt{\Delta^2-1}$ on $B_\pm$. 
The set $B_\pm$ of the boundary sites are given by 
$B_{\pm}=\{(a_x,a_y)\in\Lambda|\ a_y=\pm L_y/2\}$. 
The interaction Hamiltonian $H_{{\rm el}-{\rm dw}}$ between 
the spin ${\bf s}$ of the conduction electron and 
the localized spins ${\bf S}_{\bf a}$ is given by 
\begin{eqnarray}
H_{\rm el-dw}&=&-J_K\sum_{{\bf a}\in\Omega}
\left[s^{(1)} S_{\bf a}^{(1)}+s^{(2)} S_{\bf a}^{(2)}
+\Delta_Ks^{(3)} S_{\bf a}^{(3)}\right]\nonumber\\
& &\qquad \times\ u({\bf r}-{\bf a}),
\end{eqnarray}
where ${\bf r}=(x,y)$ is the two-dimensional coordinate of the electron, 
and $J_K$ and $\Delta_K$ are the exchange integral and the anisotropy 
of the Hund couplings, respectively. The function $u({\bf r})$ is positive 
and satisfies a short range condition 
$u({\bf r})=0$ for $|{\bf r}|>r_0$ with a positive constant $r_0\approx a$. 

\begin{figure}
\epsfxsize=80mm
\epsfbox{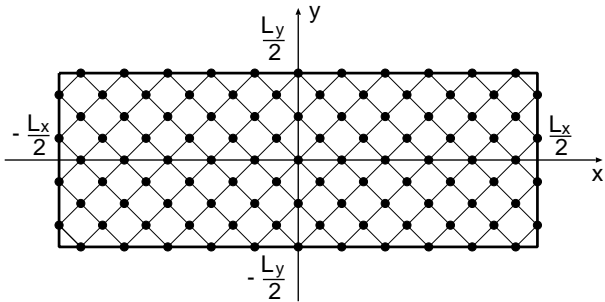}
\caption{Two dimensional diagonal lattice in the $L_x\times L_y$ box with 
the two-dimensional coordinate.} 
\label{fig:diagonalattice}
\end{figure}

The domain wall ground state \cite{REFASW,REFGW,REFmatsui,REFKN3} 
of the Hamiltonian 
$H_{\rm dw}$ is given by 
\begin{equation}
\Phi(z) = \bigotimes_{{\bf a}\in\Lambda}\eta_{\bf a}(z),
\label{eq:grandcanonicalgs}
\end{equation} 
where 
\begin{equation}
\eta_{\bf a}(z)=\frac{1}{\sqrt{1+|z|^2q^{2m}}}
\left(|\uparrow\rangle_{\bf a}+zq^m|\downarrow\rangle_{\bf a}\right) 
\label{eta}
\end{equation}
with ${\bf a}=(ma,na)$. Here $z$ is a complex number, $q$ is defined as 
$\Delta=(q + q^{-1})/2$ with $0<q<1$, 
and $|\uparrow\rangle_{\bf a}$ and $|\downarrow\rangle_{\bf a}$ are the spin 
up and down states at the lattice site ${\bf a}$, respectively. 
Write $z=e^{\ell/2+i\phi}$ with real numbers $\ell,\phi$. 
Then the position of the domain wall is specified with $\ell$ in the $x$ 
direction and the angle $\phi$ is a quantum mechanical phase corresponding to 
the degree of freedom of the rotation about the third axis of the spin. 
In the following, we choose $\ell=0$. Namely we put the center of 
the domain wall at $x=0$ in the $x$ direction. 

Recently the excitations above of the domain wall ground states in 
the ferromagnetic XXZ model with the Ising anisotropy $\Delta>1$ 
and with the spin $S\ge 1/2$ have been intensively investigated 
\cite{REFKN2,REFmatsui2,REFKN1,REFnach,REFNN}. 
Among many results, the most surprising result 
about the quantum domain walls is that, in two or higher dimensions, 
gapless excitations appear above the domain wall ground states 
\cite{REFKN2,REFmatsui2}, 
whereas, in one dimension, all of the excitations have a finite energy gap 
above the ground states \cite{REFKN2,REFKN1}. 
Here we should note that all of the excitations 
above the translationally invariant ferromagnetic ground states 
always have a finite energy gap in any dimension and for any spin $S$ 
because of the Ising anisotropy. 
In other words, the usual spin wave excitations always exhibit 
a finite energy gap. Thus the existence of the domain wall 
makes low energy excitations gapless in two or higher dimensions. 
This surprising fact was found by Koma and Nachtergaele \cite{REFKN2,REFKN1}, 
and they first proved the existence of a finite energy gap in one dimension 
and then proved the existence of a gapless excitation in two dimensions. 
The latter result was extended to three or higher dimensions 
by Matsui \cite{REFmatsui2}. 
In their mathematical proof, the gapless excitations 
were constructed by deforming the ferromagnetic domain wall 
locally \cite{REFKN2,REFmatsui2,REFnach}. 
However, in realistic situations, we expect that 
the corresponding low energy excitations appear as kinks of a domain wall. 
Namely creating a kink is the simplest deformation for the domain wall. 

We consider a single kink above the domain wall although 
there appear at least two kinks for a low energy excitation 
with a local support above the domain wall ground state. 
Namely, we study the effect of a single kink in 
the electric transport through the domain wall. 
In order to construct a single kink above the domain wall 
ground state (\ref{eq:grandcanonicalgs}), we twist 
the quantum mechanical phase along the $y$ direction, 
i.e., along the domain wall. The explicit form of the kink is 
\begin{equation}
\Phi(z;\varphi) = \bigotimes_{{\bf a}\in\Lambda}\eta_{\bf a}(z;\varphi),
\label{kinkvarphi}
\end{equation} 
with 
\begin{equation}
\eta_{\bf a}(z;\varphi)=\frac{1}{\sqrt{1+|z|^2q^{2m}}}
\left[|\uparrow\rangle_{\bf a}+zq^me^{i\varphi(na)}|\downarrow
\rangle_{\bf a}\right] 
\end{equation}
and with 
\begin{eqnarray}
\varphi(na) = \left\{\begin{array}{rl}
-\delta, & na <y_0 \\
 \delta, & na >y_0 \\
 0, & \mbox{otherwise,}
\end{array}
\right.
\end{eqnarray}
where $\delta$ is a real number, and $ y_0$ is the position 
of the kink of the domain wall. 
(In the same way, we can treat a general deformation 
$z\rightarrow z\exp[\gamma(na)+i\varphi(na)]$ with 
real functions $\gamma(na), \varphi(na)$.) In a real material, 
we can expect that the motion of the kink is much slower than 
that of the conduction electron. 
Under this assumption, the effective Hamiltonian for the conduction electron 
is given by 
\begin{eqnarray} 
{\tilde H}_{\rm eff}&=&\langle\Phi(z;\varphi), H\Phi(z;\varphi)\rangle
\nonumber\\
&=&-\frac{\hbar^2}{2\mu}\bigtriangleup
-\frac{J_K\Delta_K}{4}\tanh\Big(\frac{x}{\lambda}\Big)\sigma^{(3)}\nonumber\\
& &-\frac{J_K}{4}{\rm sech}\Big(\frac{x}{\lambda}\Big)
\exp\left[i\{\varphi(y)+\phi\}\sigma^{(3)}\right]\sigma^{(1)}\nonumber\\
& &+{\rm const.},
\label{tildeHeff}
\end{eqnarray} 
where $(\sigma^{(1)},\sigma^{(2)},\sigma^{(3)})$ is the Pauli matrix, 
the width $\lambda$ of the domain wall is given by $\lambda=a/|\log q|$, 
and we have replaced $(ma,na)$ with $(x,y)$ and chosen 
$u({\bf r})\approx 1$ in the interaction range 
because we can expect that the qualitative scattering behavior 
of the conduction electron by the effective domain wall potential 
is independent of the detail of the lattice structure. 

Let us consider a wire with the width $L_y$ in the $y$ direction 
and with the infinitely long length $L_x=+\infty$ in the $x$ direction. 
We impose the Dirichlet boundary conditions $\psi(x,\pm L_y/2)=0$ 
in the $y$ direction for the wavefunction $\psi({\bf r})$ of the electron. 
Since the scattering potential of the domain wall depends on 
the position $y_0$ of the kink, we can expect that 
the transmission coefficient $T$ through the domain wall potential 
varies depending on the position $y_0$. 
In order to see the dependence explicitly, we shall introduce 
an approximation. For this purpose, consider first the case 
with no domain wall, i.e., no effective domain wall potential 
in the effective Hamiltonian (\ref{tildeHeff}). 
Then the wavefunction $\psi({\bf r})$ of the electron has a product 
form $\psi({\bf r})=\psi_x(x)\psi_y(y)$ 
because the Hamiltonian ${\tilde H}_{\rm eff}$ of (\ref{tildeHeff}) is exactly 
of the free electron form. Now our approximation is as follows: 
We restrict the wavefunction $\psi_y(y)$ 
to the sector of the ground state 
\begin{equation}
\psi_y^{(0)}(y)=\sqrt{\frac{2}{L_y}}\cos\frac{\pi}{L_y}y.
\end{equation}
Namely we ignore all the excitations from the lowest subband to 
the higher subbands. Although this approximation is not necessarily 
justified in a realistic situation, we believe that a similar position 
dependence inevitably appears in the transmission coefficient $T$. 
With the approximation, the effective Hamiltonian $H_{\rm eff}$ for 
the conduction electron is given by 
\begin{eqnarray} 
H_{\rm eff}&=&\langle\psi_y^{(0)},{\tilde H}_{\rm eff}\psi_y^{(0)}\rangle
\nonumber\\
&=&-\frac{\hbar^2}{2\mu}\frac{\partial^2}{\partial x^2}
-\frac{J_K\Delta_K}{4}\tanh\Big(\frac{x}{\lambda}\Big)\sigma^{(3)}\nonumber\\
& &-\frac{J_K}{4}{\rm sech}\Big(\frac{x}{\lambda}\Big)
A(y_0,L_y)\sigma^{(1)}+{\rm const.},
\label{Heff}
\end{eqnarray} 
where 
\begin{equation}
A(y_0,L_y)=\sqrt{\cos^2\delta+\left(\frac{2y_0}{L_y}+
\frac{1}{\pi}\sin\frac{2\pi}{L_y}y_0\right)^2\sin^2\delta}
\end{equation}
and we have chosen $\phi$ in $z$ as 
\begin{equation}
\phi=\tan^{-1}\left[\left(\frac{2y_0}{L_y}+
\frac{1}{\pi}\sin\frac{2\pi}{L_y}y_0\right)^2\tan\delta
\right].
\end{equation}
For $\delta=0$, we recover the well-known effective Hamiltonian 
\cite{REFYK,CF} for the conduction electron with the single domain wall 
with no kink. Clearly the transmission coefficient $T$ is a function 
of $A(y_0,L_y)$. Let us consider the case with a small $\delta$, 
i.e., with a very low energy excitation above the domain wall. 
We expand $T=T(A(y_0,L_y))$ as 
\begin{eqnarray}
T&=&T_0+
{\rm const.}\times\left\{\cos^2\delta-1+\left[f(y_0,L_y)\right]^2\sin^2\delta
\right\}\nonumber\\
&+&{\cal O}(\delta^4),
\label{Tresult1}
\end{eqnarray}
where 
\begin{equation}
f(y_0,L_y)=
\frac{2y_0}{L_y}+\frac{1}{\pi}\sin\frac{2\pi}{L_y}y_0,
\end{equation}
and $T_0$ is the transmission coefficient in the case with no kink. 
The second term varies as the position $y_0$ of the kink 
varies from $-L_y/2$ to $L_y/2$. (See Fig.~2.) 
Thus the conductance which is proportional to the transmission 
coefficient $T$ varies depending on the position $y_0$ of the kink 
when the kink moves on the domain wall, 
owing to the thermal fluctuations, or to the driving force of 
an external magnetic field. In other words, when the position $y_0$ of the 
kink dynamically fluctuates, the observed resistance also fluctuates 
reflecting the fluctuations of the kink.

\begin{figure}
\epsfxsize=80mm
\epsfbox{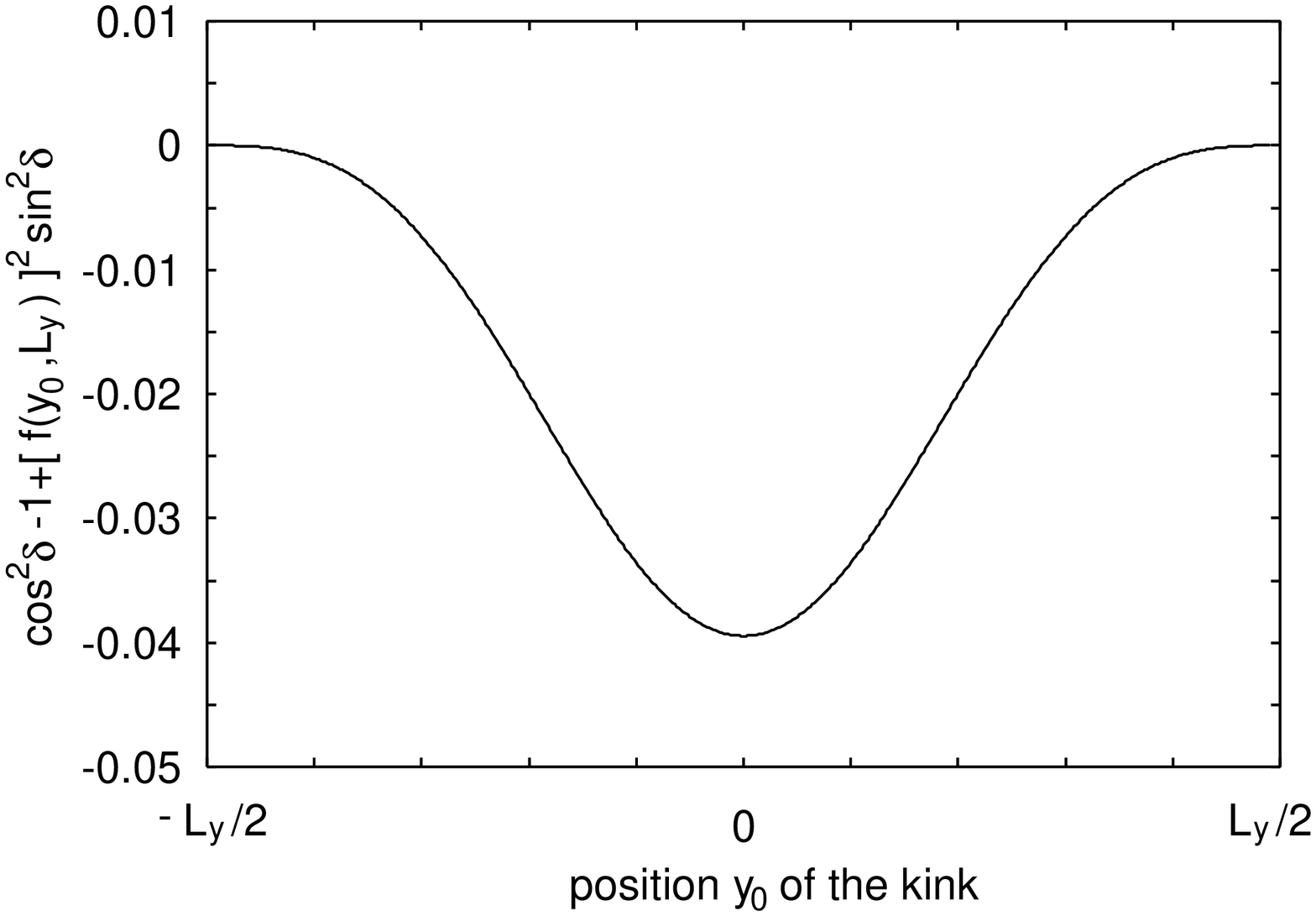}
\smallskip\smallskip\smallskip\smallskip
\caption{The quantity 
$\cos^2\delta$$-1$$+\left[f(y_0,L_y)\right]^2\sin^2\delta$
in Eq.~(15) as a function of the position $y_0$ of the kink for $\delta=0.2$.
The effect due to the kink in the scattering of the conduction electron 
is maximized when the kink is at the center of the sample 
in the $y$ direction.} 
\label{fig:function}
\end{figure}

Let us briefly discuss a possibility of detecting the gapless mode 
in neutron scattering or ferromagnetic resonance experiments. 
The dominant part of their signals consists of the contributions 
from the usual ferromagnetic spin wave excitations which have a finite 
energy gap. The spectrum of the gapless mode appears below those of 
the spin wave excitations. 
The signal from the gapless mode is expected to be very weak in comparison 
with those from the spin wave modes because 
the spin wave excitations move around the whole crystal, 
whereas the gapless mode is confined in the narrow regions 
along the domain walls. An effective theory to describe the gapless mode 
was given in Ref.~\cite{REFnach}. 

Using neutron scattering, the Fourier transforms of the spin-spin correlation 
functions can be obtained experimentally. The results also include 
the information of the energy-momentum relations for the usual spin 
wave excitations and for the gapless excitations. 
However, the intensity of the signal from the gapless mode 
is expected to be very weak as mentioned above. 
Recently the dynamical properties of the spin-spin correlation 
for a domain wall state were studied theoretically, and the contribution 
from the gapless mode was obtained within a random phase 
approximation \cite{REFNN}. 
We can expect that the detection of the gapless mode in a neutron scattering 
experiment will succeed by overcoming the technical problem 
in the near future, and that the experimental results will be compared 
to some theoretical results. 

We also expect that a ferromagnetic resonance experiment is useful 
for measuring the spectrum of the gapless mode. 
In addition, such an experiment generally have the advantage 
that the observed signal profile strongly depends on the geometry 
of a sample and an external magnetic field. 
These properties may be exploited for detecting the gapless mode. 
For example, consider a three-dimensional anisotropic 
$L_x\times L_y\times L_z$ system satisfying $L_z\ll L_y\ll L_x$, 
i.e., a long wire with the anisotropic widths $L_y, L_z$. 
For small widths $L_y, L_z$, the usual spin waves in the $y$ and 
$z$ directions become standing waves with a discrete spectrum because of 
the finite size effect \cite{REFwalker}. 
Further, if we take the width $L_z$ sufficiently small, 
the energies of the standing waves in the $z$ direction 
become much higher than the rest of the spectrum. 
In this situation, we can ignore these modes in the $z$ direction. 
Assume that the face of a single domain wall entered the wire 
is perpendicular to the longitudinal $x$ direction. 
Then we can expect that the signal from the excitation modes 
in the $y$ direction is different from that without domain walls 
because there appear peculiar excitations, such as a kink 
excitation, due to the presence of the domain wall. 
Namely we can expect that the spectrum of the gapless mode 
is observed below that of the usual spin waves, 
by controlling the width $L_y$ of the wire. 

In a ferromagnetic resonance experiment, 
one need to take the frequency of the applied microwaves to be very low 
because the energies of the gapless mode are very low. 
In order to estimate the magnitude of the frequency, consider 
the sample in Ref.~\cite{Otani} as an example. 
Then the external magnetic field $\mu_0 H$ to create a domain wall 
into the sample is of order of $10$ mT. 
The corresponding frequency of the microwave is estimated 
as $1\sim100$ MHz or less. Therefore the natural ferromagnetic resonance 
in the absence of the static magnetic field
or using the terrestial magnetic field seems to be suitable for detecting 
gapless mode above ferromagnetic domain walls. 
The oscillating magnetic field of the applied microwave is set
to be either perpendicular or parallel to the internal 
or terrestial magnetic field. 
This argument about the estimate for the frequency of the microwaves 
is based on a discussion with K.~Furukawa and T.~Takui \cite{REFfurukawa}. 

The ferromagnetic resonance for a sample including point-like 
magnetic objects, such as localized magnetic impurities, 
has been often investigated so far. 
However, as far as we know, a system including extended magnetic 
objects with an internal degree of freedom, such as domain walls with 
kinks, has not yet been investigated. 

We are grateful to Kou Furukawa and Takeji Takui
for a stimulating discussion on ferromagnetic resonance experiments. 
We also would like to thank 
Shinji Nonoyama and Yoshichika Otani for many useful discussions.
M.Y. is supported by the Moritani Scholarship Foundation.


\begin{references}
\bibitem{diagonal} In order to create a domain wall along the $y$ axis, 
we need to take the lattice to be diagonal.
\end{references}
\end{document}